%% file: main.tex
\documentclass[journal=jctcce, manuscript=article, layout=onecolumn]{achemso}
\setkeys{acs}{articletitle = true}
\usepackage{amsmath}

\input{command}

\usepackage{ulem}

\title{Machine Learning Framework for Modeling Exciton-Polaritons in Molecular Materials}

\author{Xinyang Li}
\affiliation{Physics and Chemistry of Materials, Theoretical Division, Los Alamos National Laboratory, Los Alamos, New Mexico, 87545, USA}

\author{Nicholas Lubbers}
\affiliation{Information Sciences, Computer, Computational, and Statistical Sciences Division, Los Alamos National Laboratory, Los Alamos, NM 87545, USA}

\author{Sergei Tretiak}
\affiliation{Physics and Chemistry of Materials, Theoretical Division, Los Alamos National Laboratory, Los Alamos, New Mexico, 87545, USA}
\alsoaffiliation{Center for Integrated Nanotechnologies, Los Alamos National Laboratory, Los Alamos, New Mexico 87545, USA}

\author{Kipton Barros}
\email{kbarros@lanl.gov}
\affiliation{Physics and Chemistry of Materials, Theoretical Division, Los Alamos National Laboratory, Los Alamos, New Mexico, 87545, USA}

\author{Yu Zhang}
\email{zhy@lanl.gov}
\affiliation{Physics and Chemistry of Materials, Theoretical Division, Los Alamos National Laboratory, Los Alamos, New Mexico, 87545, USA}

\date{\today}

\begin{document}

\begin{tocentry}
  \includegraphics[width=\textwidth]{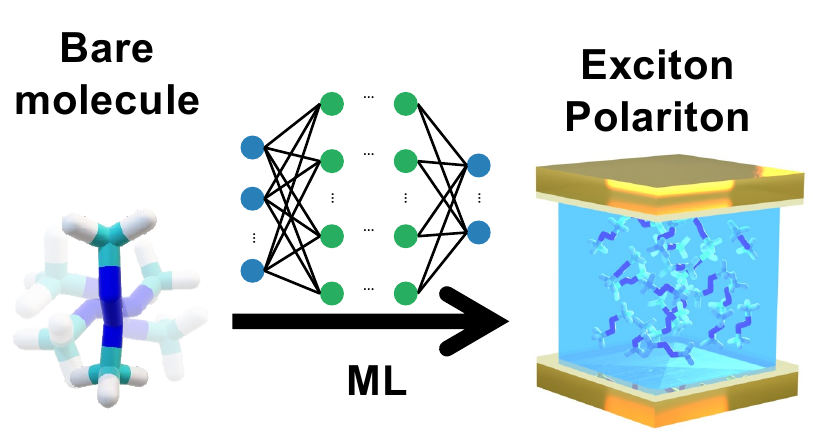}
\end{tocentry}

\begin{abstract}
A light-matter hybrid quasiparticle, called a polariton, is formed when molecules are strongly coupled to an optical cavity. 
Recent experiments have shown that polariton chemistry can manipulate chemical reactions. Polariton chemistry is a collective phenomenon and its effects increase with the number of molecules in a cavity. 
However, simulating an ensemble of molecules in the excited state coupled to a cavity mode is theoretically and computationally challenging. Recent advances in machine learning techniques have shown promising capabilities in modeling ground state chemical systems.
This work presents a general protocol to predict excited-state properties, such as energies, transition dipoles, and non-adiabatic coupling vectors with the hierarchically interacting particle neural network.
Machine learning predictions are then applied to compute potential energy surfaces and electronic spectra of a prototype azomethane molecule in the collective coupling scenario.
These computational tools provide a much-needed framework to model and understand many molecules' emerging excited-state polariton chemistry.
\end{abstract}

\maketitle

\section{Introduction}\label{seq:intro}
When excitations in matter hybridize with the photonic excitations, a new quasiparticle, the polariton, is formed.~\cite{Dicke1954, Hopfield1958, Ritsch2013, Deng2010RMP}
Consequently, when molecular excitations, such as electronic excitations, hybridize with photonic excitations, this creates molecular polaritons.~\cite{Hutchison2012ACIE, GarciaVidal2021}
Even though polaritons have been known in physics for decades, only in recent years there have been significant advancements in using molecular polaritons to tune the excited-state potential energy surface (PES) of molecules.
Thus, molecular polaritons provide an alternative and attractive way to manipulate the chemical dynamics,~\cite{Ebbesen2016ACR, Sanvitto2016NM, Mandal2022} including long-range energy transfer,~\cite{Zhong2017ACIE,Du2018CS,Saez-Blazquez2018} enhanced charge transfer,~\cite{Orgiu2015NM, Mandal2020, DelPo2021} and polariton lasing,~\cite{Ramezani2016O, KenaCohen2010NP, Rajendran2019} and organic condensate.~\cite{Kavokin:2022tn, Dusel:2020wt, Deng2010RMP}
The groundbreaking experiments conducted by the group of Thomas Ebbesen showed that strong coupling
could affect the PES landscape, which then alters the rate of photochemical
reactions.~\cite{Hutchison2012ACIE} This possibility has inspired the appearance of polaritonic
chemistry aiming to manipulate chemical structures and reactions via the formation of polaritons,
which has become a topic of intense experimental~\cite{Wang2014N, Zeng2016NL, Baieva2017AP} and
theoretical research~\cite{MartinezMartinez2017AP, Flick2017JCTC, Zeb2017AP, Galego2016NC,
Barachati2017AP, Berenbeim2017JPCL, Mandal2022, Weight2023,Weight2023a,Li2023} in the past few years. In addition, recent developments have
found that vibrational strong coupling (VSC) can resonantly enhance or
suppress~\cite{CamposGonzalezAngulo2019} thermally-activated chemical reactions and change selectivity
between competing reactions~\cite{Thomas2019S, Li2021JPCL} via the formation of vibrational
polaritons. 

However, modeling molecular polaritons theoretically is a non-trivial problem. Polariton chemistry is inherently a many-molecule effect, often involving collective coupling of over $10^6$ molecules to the cavity mode.~\cite{Ebbesen2016ACR, Pino2015, CamposGonzalezAngulo2019}
A quantitative quantum model accounting for numerous organic molecules strongly coupled to a confined light mode must encompass principles from both quantum electrodynamics (QED) and quantum chemistry.~\cite{Galego2015PRX, Herrera2016PRL, Feist2017AP, Flick2017PNAS, Flick2015PNAS} Even though QED itself has been established for several decades,~\cite{Feynman1988} its integration with quantum chemistry is still not well understood.
In addition, excited-state polariton chemistry involves a complex interplay among the nuclear, electronic, and photonic degrees of freedom at different length and time scales.
As such, a theory incorporating non-adiabatic effects is required to describe photochemistry under the strong light-matter interaction, where molecular dynamics of electronically excited states are essential. While rigorous methods accounting for non-adiabatic quantum dynamics have been proposed for strong light-matter interaction, the computational complexity of quantum dynamics ultimately limits its applications to minimal systems where only a few PESs and dimensions can be treated.~\cite{Kowalewski2016JPCL, Kowalewski2016JCP, Bennett2016FD, Kowalewski2017PNAS, Triana2018JPCA, Mandal2019JPCL} Consequently, due to this theoretical challenge and computational cost, most prior theoretical work has been limited to single-molecule or few-molecule level of treatment, which is far from the experimental setups.~\cite{} To address this issue, mixed quantum-classical methods, such as Ehrenfest dynamics and trajectory surface hopping (TSH),~\cite{Tully2012} may offer an alternative approach for modeling polariton chemistry.
However, it remains a challenge to extend these methods to a large number of molecules.

Alternatively, machine learning (ML) methods, especially deep neural networks (NN), are emerging techniques that have been extensively used to study molecular systems in the ground state, especially molecular energies.~\cite{Behler2007PRL, Rupp2012PRL, Montavon2012NIPS, Bartok2013PRB, Lilienfeld2015IJQC,
Shapeev2016MMS, De2016PCCP, Huo2022MLST, Behler2015IJQC, Faber2017JCTC, Artrith2017PRB,
Hansen2015JPCL, Gubaev2018JCP, Rupp2015JPCL}
The most notable feature of this approach is that it achieves a commendable balance between accuracy and computational cost. Upon training on appropriate datasets, the accuracy of the predictions may be comparable to that of the density functional theory~\cite{Faber2017JCTC} or even the coupled cluster theory~\cite{Smith2019NC}, while maintaining a computational cost akin to the classical force fields. 
Consequently, integrating NNs with mixed quantum-classical methods can be a promising approach for simulating complex systems like exciton polaritons. 
This setup allows us to scale
simulations up to thousands of or even more molecules coupled to the cavity mode.
A number of previous studies on utilizing ML to predict excited-state quantities have been published recently.~\cite{Westermayr2020CR} These works include predicting adiabatic excited-state PES with NN~\cite{Guan2017} or traditional ML methods, like kernel ridge regression (KRR)~\cite{Hu2018, Dral2018} or Gaussian process regression (GPR),~\cite{Richings2017} transition dipole modeled with KRR~\cite{Hu2023}, dipole moments and PES in diabatic representation,~\cite{Guan2019, Guan2020} non-adiabatic coupling vectors,~\cite{Westermayr2019CS, Richardson2023} and attempts on achieving transferability for excited-state modeling.~\cite{Westermayr2020JCP}
Normally, a non-adiabatic simulation involves a manifold of excited states, with frequent crossings among them. The characteristics and identity of an adiabatic state frequently change in the course of these crossings. This dynamical process is well understood and addressed in conventional non-adiabatic simulations using various algorithms.~\cite{Tully2012, Malone2020}
This issue can still affect the training of ML models, making predictions more challenging. Further, to accurately calculate transition rates between excited states in the presence of light-matter coupling, reasonably accurate predictions of excitation energies and their gradients, transition dipole moments, and non-adiabatic coupling vectors (NACRs) are essential, in contrast, to ground-state adiabatic dynamics which requires only the ground-state energy and its gradients (forces).

In this work, we present a general protocol to evaluate the excited-state properties, such as energies, transition dipole moments, and NACRs, for application in computing the polaritonic eigenstates of many molecules in cavities, with a single Hierarchically Interacting Particle Neural Network (HIP-NN).~\cite{Lubbers2018JCP, Chigaev2023} A new way for representing NACRs with transition atomic charges is proposed. We validate the predictive accuracy of this method using a NN model trained on the azomethane molecule. Further, we use these predictions to compute and analyze the polariton energies and spectra in various scenarios.
It should be noted that the molecules in a polariton chemistry setup are the same but in different configurations due to thermal fluctuations and transient dynamics.
As a result, we only need to train an NN model predicting energies, dipoles, and NACRs for one molecule and apply the NN model to compute the polariton eigenstates of many molecules, as shown in Sec.~\ref{sec:pred}.
Though intermolecular interactions are not explicitly considered in the one-molecule NN model, the intermolecular interaction can be approximated by the dipole-dipole interactions when applying the model to study the many-molecule cases.
Consequently, our NN model bypasses the need for resource-intensive electronic structure calculations of many molecules explicitly.

\section{Methods}

\subsection{Quantum electrodynamics Hamiltonians}

To provide a comprehensive description of the light-matter coupled system, the quantum optics community has proposed a number of theoretical frameworks featuring distinct Hamiltonians.
Two commonly used ones are the Jaynes-Cummings (JC) model~\cite{Jaynes1963} and Tavis-Cummings (TC) model.~\cite{Tavis1968,Tavis1969}

The JC model considers a two-level system interacting with a lossless cavity mode under the rotating wave approximation (RWA),~\cite{Scully2012} which is described by the following Hamiltonian,

\begin{align}
  \label{eq:jc}
  \hat{H}_{JC} = \omega_c a^\dagger a +
    \Omega(\boldsymbol{R})\sigma^\dagger \sigma + g_c(\boldsymbol{R})(a^\dagger \sigma + a \sigma^\dagger),
\end{align}
where the molecule has two states, the ground state (S$_0$) and the excited state (S$_1$).
$\Omega(\boldsymbol{R})$ is the energy gap between S$_0$ and S$_1$ along the reaction coordinate
$\boldsymbol{R}$. $a^\dagger$($a$) is the photon creation (annihilation) operator. $\omega_c$ is the
cavity photon frequency. $\sigma^\dagger = \ket{S_1}\bra{S_0}$ and $\sigma = \ket{S_0}\bra{S_1}$ are
the creation and annihilation operators for molecular excitation, respectively.

The third term in Eq.~\ref{eq:jc} represents the light-matter coupling term. $g_c$ characterizes the
light-matter coupling strength, which can be expressed as,
\begin{align}
  \label{eq:gc_dot}
  g_c = \boldsymbol{\mu}(\boldsymbol{R}) \cdot \boldsymbol{\mathcal{E}},
\end{align}
where $\boldsymbol{\mu}(\boldsymbol{R})$ is the transition dipole moment of the molecule between
S$_0$ and S$_1$. $\boldsymbol{\mathcal{E}}$ is the vector of the cavity electric field. If the
transition dipole is always assumed to be aligned with respect to the electric field, the coupling
strength can be simplified as
\begin{align}
  g_c = g \lVert \boldsymbol{\mu} \rVert,
\end{align}
where $g$ is a user-defined parameter that measures the coupling strength.

The JC model can easily be generalized to include multiple excited states,~\cite{Zhang2019}
\begin{align}
  \label{eq:jc_N}
  \hat{H}_{pl} = \omega_c a^\dagger a + \sum_{i=1}^{n} \Omega_i(\boldsymbol{R}) 
    \sigma_i^\dagger \sigma_i + g_c^i(a^\dagger \sigma_i + a \sigma_i^\dagger),
\end{align}
where the index $i$ represents the $i^\mathrm{th}$ excited state and $n$ the total number of excited state considered.

TC model, on the other hand, is used to model $N$ molecules of the same species coupled to a lossless cavity mode, i.e., it is a many-molecule extension of the JC model. 
If we further ignore the intermolecular interactions, the expression of the TC model is identical to Eq.~\eqref{eq:jc_N}, except that the index $i$ refers to a molecule instead of a state.

\begin{figure*}
  \centering
  \includegraphics[width=0.49\textwidth]{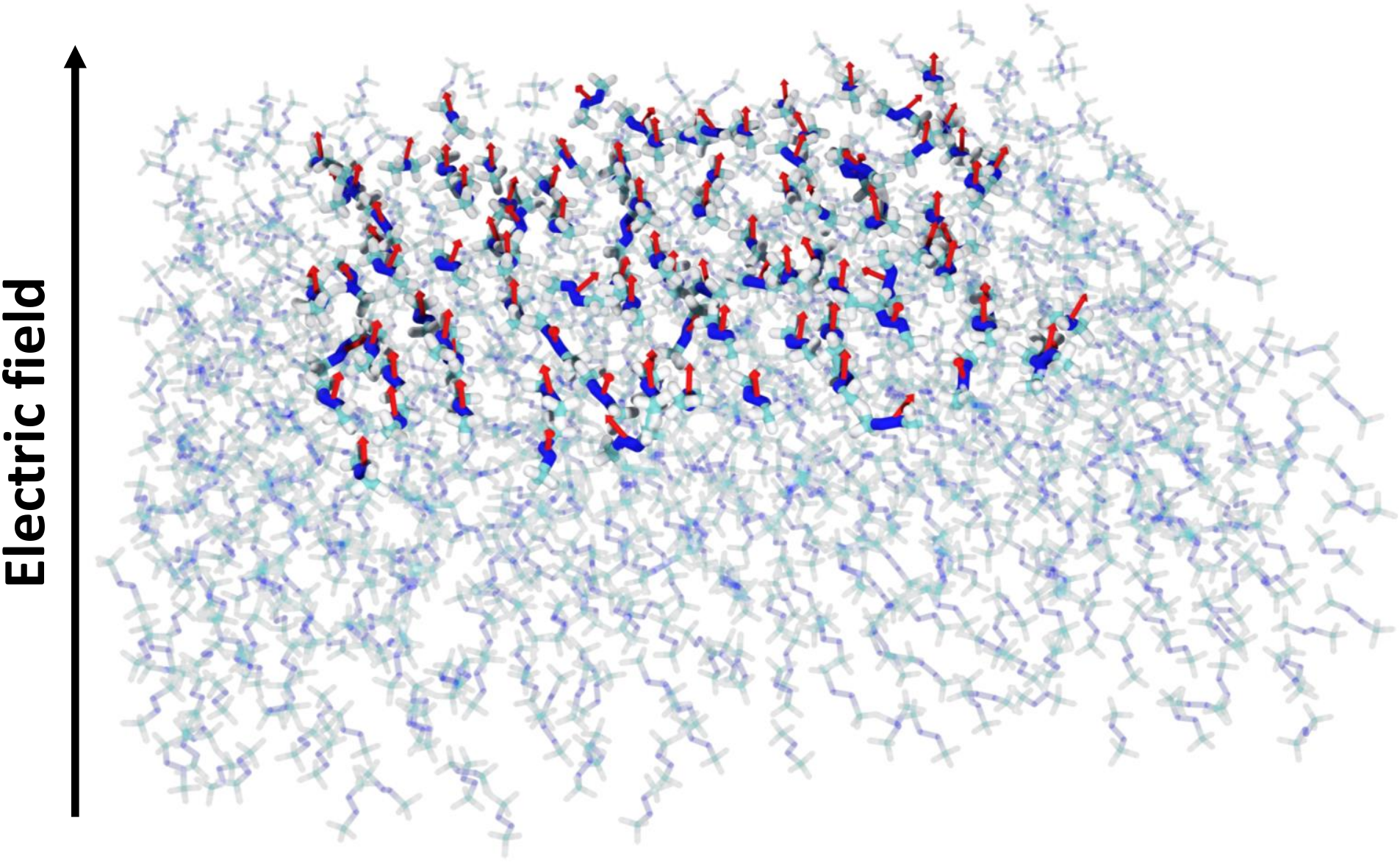}
  \caption{Schematic illustration of 1,000 molecules coupled to a cavity mode. Red arrows depict the directions of the transition dipoles of the highlighted molecules. The angles between the transition dipoles and the electric field are modeled by a Gaussian function centered around 0\degree~with a full width at half maximum (FWHM) of 60\degree.}
  \label{fig:model}
\end{figure*}

\input{hipnn_diagram}

The light-matter coupled system is illustrated in Figure~\ref{fig:model}, where 1,000 molecules with different geometries and orientations are coupled to the cavity.
Here, we assume that due to the strong interaction between the molecule and the field, molecules will have the tendency to align along the direction of the electric field.
However, because of thermal fluctuations, they will not always align perfectly.
As a result, we use a Gaussian function centered around 0\degree~ to model this disorder.
This model is used throughout this work.
It is essential to emphasize that this constitutes a qualitative description of a light-matter coupled system. A quantitative analysis of such disorder necessitates MD relaxation of all the molecules within realistic cavities, which will be the focus of future studies. 

\subsection{Hierarchically interacting particle neural network (HIP-NN)}
\label{sec:hip-nn}

Here, we provide a concise overview of the HIP-NN model. For additional details, we refer readers to Ref.~\citenum{Lubbers2018JCP}. HIP-NN can be interpreted as a message-passing neural network, which updates atomic features based on the information from nearby atoms, or as a continuous convolutional neural network, processing atoms by the local distribution of features in space.
This network translates atomic coordinates into pairwise interatomic distances, which serve as inputs for the HIP-NN, ensuring its invariance to symmetry operations such as translations, rotations, and permutations.
It integrates a locality assumption, such that individual operations in the network only take into account distances $r_{ij} < R_{\mathrm{cut}}$, where $R_\mathrm{cut}$ is a user-defined cutoff radius.

Figure~\ref{fig:hipnn} shows the data processing procedure within the network.
The HIP-NN architecture consists of two types of layers: on-site layers that locally process atomic features (red squares in Figure~\ref{fig:hipnn}) and interaction layers that exchange information among neighboring atoms (green rectangles in Figure~\ref{fig:hipnn}).
The on-site layers are fully connected and activated by a softplus function.

In contrast to the on-site layers, the interaction layers incorporate the interactions between an atom and its neighboring atoms within the user-specified cutoff distance $R_\mathrm{cut}$, thereby capturing non-local contributions.
To capture interatomic interactions, a linear combination of a basis of sensitivity functions is used. These sensitivity functions are Gaussian functions in inverse pairwise distances governed by the cutoff distance $R_{cut}$. The number of sensitivity functions used is a user-defined parameter, which requires hyper-parameter optimizations.

The layers in HIP-NN are organized into hierarchical blocks. Each block begins with an interaction layer followed by several on-site layers, as shown in Figure~\ref{fig:hipnn}.
Atomic properties (atomic energy $E_i$, for example) are predicted hierarchically by summing up the predictions from the last layer $l_n$ of each hierarchical block $n$,
\begin{align}
    \tilde{E}_i = \sum_n \tilde{E}^{(n)}_i = \sum^{N_\mathrm{features}}_{a=1} \omega^{n}_{a} z^{l_{n}}_{i,a} + b^{n}.
\end{align}
Note that terms at higher hierarchical order $n$ can include interaction from further neighbors. The effective cutoff distance ("receptive field") is $n \cdot R_\mathrm{cut}$. The molecular quantity (energy, for example) is a sum of all the atomic contributions,
\begin{align}
  \tilde{E} = \sum_{i=1}^{N_{atom}} \tilde{E}_i,
\end{align}
where the ``tilde'' denotes predicted values. The atomic transition charges $\tilde{\boldsymbol{q}}_i$ are predicted in the same fashion. The predicted transitions dipoles $\tilde{\boldsymbol{\mu}}_i$ and NACRs $\tilde{\boldsymbol{d}}_{ij}$ are derived from the charges with details provided in the following section.

\subsection{Electronic structure and the neural network treatment}
\label{subsec:ee}

The molecular energy of the $i^\mathrm{th}$ state is expressed as
\begin{align}
  E_i = \braket{ \Psi_i({\boldsymbol{R}}) | \hat{\boldsymbol H} | \Psi_i ({\boldsymbol{R}}) },
\end{align}
where $\Psi_i({\boldsymbol{R}})$ is the wavefunction of state $i$ and $\hat{\boldsymbol H}$ is the
Hamiltonian. The same approach can be used to predict both ground-state and excited-state energies.

The dipole operator of a molecule can be written as
\begin{align}
  \hat{\boldsymbol \mu}=\sum_k z_k \hat{\boldsymbol{x}}_k,
\end{align}
where $z_k$ is the charge for the $k^\mathrm{th}$ charged particle. The transition dipole moment can be calculated using the adiabatic states,
\begin{align}
    {\boldsymbol \mu}_{i} ({\boldsymbol{R}}) = \langle \Psi_0({\boldsymbol{R}}) | \hat{\boldsymbol\mu} | \Psi_i ({\boldsymbol{R}}) \rangle,
\end{align}
where $\Psi_0({\boldsymbol{R}})$ is the ground-state wavefunction.

To predict the transition dipole with NNs, we decompose the total dipole moment into a linear combination of atomic dipoles, which can be related to atomic transition charges.
\begin{align}
    {\boldsymbol \mu}_{i} ({\boldsymbol{R}}) = \sum_k {\boldsymbol \mu}_{i, k}
    = \sum_k q_{i, k} \boldsymbol{R}_k,
\end{align}
where $q_{i, k}$ is the transition atomic charge for the $k^\mathrm{th}$ atom from the ground state to the $i^\mathrm{th}$ excited state. In the HIP-NN framework, $\{ q_{i, k} \}$ are latent variables inferred from the dipole fitting and not directly exposed to users.

The non-Born-Oppenheimer effects and non-adiabatic transitions between different electronic states are determined by the so-called non-adiabatic couplings (NACs). The non-adiabatic coupling vector (NACR) is defined as
\begin{align}
  \boldsymbol{d}_{ij} = \braket{\Psi_i | \nabla_{\boldsymbol{R}} \Psi_j}.
\end{align}
The state transition probability can be related to the non-adiabatic coupling term (NACT),~\cite{Malone2020}
\begin{align}
  |d_{ij}| \equiv |\boldsymbol{d}_{ij} \cdot \boldsymbol{v}|,
\end{align}
where $\boldsymbol{v}$ is the nuclear velocities. As NACT is a time-dependent variable (due to velocities), we cannot directly predict this quantity based on the geometries.
As a result, we will instead predict the NACR vector.

The NACR can be written in the Hellmann-Feynman-like form as
\begin{align}
  \boldsymbol{d}_{ij} = \braket{\Psi_i | \nabla_{\boldsymbol{R}} \Psi_j}
  = \frac{\braket{\Psi_i | \nabla_{\boldsymbol{R}}\hat{\boldsymbol{H}} | \Psi_j}}{E_j - E_i}.
\end{align}
However, this expression shows that the NACR elements become singular near the state intersections, $E_j - E_i \to 0$.
This becomes problematic for the NNs to make predictions.
Instead, we train the NN with respect to $\boldsymbol{d}_{ij}(E_j - E_i)$, which is a well-defined quantity.~\cite{Sifain2021,Westermayr2020CR, Westermayr2020JPCL, Richardson2023}

In the current implementation, we use the transition atomic charges $\{\boldsymbol{q}_i\}$ to express NACR ($\boldsymbol{d}_{ij}$). Here the transition atomic charges on each atom are readily obtained from the adiabatic wavefunction $q^A_i(\boldsymbol{r})=\int_{\boldsymbol{r}\in A}\Psi^2_i(\boldsymbol{r})d\boldsymbol{r}$. The total wavefunction for state $i$ can be written as a summation of the ground-state wavefunction and a state-dependent perturbation, $\Psi_i = \Psi_0 + \Psi'_i$, where $\Psi_0$ is the ground-state wavefunction. As a result, we can express the state-dependent part as a functional of atomic charge density,
\begin{align}
  \label{eq:psi-q}
  \Psi_i = \mathcal{F}(\boldsymbol{q}_i),
\end{align}
where $\mathcal{F}$ represents the unknown form of the functional.
Therefore, we can express the NACR vector with transition atomic charges as,
\begin{align}
  \boldsymbol{d}_{ij} = \braket{\Psi_i | \nabla_{\boldsymbol{R}} \Psi_j}
  \sim \braket{\Psi'_i | \nabla_{\boldsymbol{R}} \Psi'_j}
  = \boldsymbol{C}_{ij}\boldsymbol{q}_i
    \frac{\partial\boldsymbol{q}_j}{\partial\boldsymbol{R}},
\end{align}
where $\boldsymbol{C}_{ij}$ is a learnable tensor that arises from Eq.~\ref{eq:psi-q}.
Note that the charge tensors for predicting NACR do not have to be the same as those for dipole predictions. The charges in this equation represent an ansatz for the neural network to use site-based quantities to model the NACR.

\subsection{Phase-less loss function}
\label{subsec:phaseless}

Wavefunctions $\{\Psi_i\}$ are defined up to a phase factor and thus have arbitrary signs. As a result, quantities that are related to two
different states, such as NACRs and transition dipoles, will also have arbitrary signs. This sign problem will cause NN predictions to collapse to 0, as the same input (geometry) can lead to opposite signs of outputs. 
To mitigate this numerical problem, we use the phase-less loss function for these quantities.~\cite{Westermayr2020CR} The phase-less version of the mean absolute error
(MAE) can be expressed as
\begin{align}
  \mathrm{MAE}'(\tilde{y},y) = \min(|\tilde{y} - y|, |\tilde{y} + y|),
\end{align}
where $\tilde{y}$ represents the predicted value, $y$ represents the training data, and $|\dots|$ refers to the L1 norm (the sum of the absolute values in the vector). When the predicted quantity is a vector, e.g., the transition dipole, the plus or minus sign is applied to the entire vector. As such, the direction of the whole vector is flipped (not individual components). Note that this approach does result in completely random signs for dipoles and NACR in static calculations, a common situation encountered in an arbitrary single-point calculation. However, when the trained model is used in molecular dynamics simulations, the signs must be tracked and corrected at every time step to avoid a sudden change in phase, following the same procedure as in conventional non-adiabatic molecular dynamics simulations~\cite{zhang2020jctc}.

The phase-less version of the root mean squared error (RMSE) is implemented in a similar way
\begin{align}
  \mathrm{RMSE}'(\tilde{y},y) = \min(\lVert \tilde{y} - y \rVert_2, \lVert \tilde{y} + y \rVert_2),
\end{align}
where the L2 norm (the Euclidean norm) is used.
The idea is that for a quantity with a phase uncertainty, as long as the prediction itself ($\tilde{y}$) or its opposite ($-\tilde{y}$) agrees with the true value ($y$), we consider it as a good prediction.
As a result, we test both $\pm\tilde{y}$ against the true value $y$, and assume that the one with a smaller error has a ``correct'' sign, and this error is then used in the loss functions. Note that this approach only ensures that the network predictions are as close as possible to either $\pm y$.
Following this procedure, the predicted values retain random signs. As discussed above, the signs are tracked and corrected in dynamics simulations.

All operations in Section~\ref{subsec:ee} and
Section~\ref{subsec:phaseless} are implemented in the open-source Python package \texttt{hippynn}.~\cite{hippynn}

\section{Results and Discussions}

\subsection{Data preparation}
\label{sec:dataset}

We chose the azomethane molecule to demonstrate the implementation of the approach described above. 
The training data is generated with the open source NEXMD package~\cite{Malone2020,Nelson2020,Freixas2023} following the common non-adiabatic molecular dynamics (NAMD) procedure.
The semi-empirical Austin Model 1 (AM1)~\cite{Dewar1985}, and configuration interaction singles (CIS) are chosen for the electronic structure simulations.
Trajectory surface hopping (TSH) is used for the NAMD simulations.~\cite{Tully2012}
The first three excited states of azomethane are included in the simulations. The snapshots are collected every 1 femtosecond (fs) or when there is a strong coupling between any two states.
The existence of strong coupling is determined by NACT. 
Specifically, when the absolute value of NACT is larger than a threshold, the snapshot is added to the dataset,
\begin{align}
  |d_{ij}| > \delta,
\end{align}
where $\delta$ is the user-defined threshold. Here, the threshold is chosen to  $1 \times 10^{-3}$ 1/ps.
$\hbar d_{ij}$ can be viewed as the energy fluctuations on the excited states due to coupling to nuclei. Any fluctuation larger than $\hbar d_{ij}$ is added to the dataset. As such, the respective threshold value must be chosen to balance between accuracy and efficiency. When its value is too small, the dataset will become very large, since it will contain many snapshots that are very close to each other, making the training slow. On the other hand, if the threshold is too large, some essential parts of the PES will be excluded from the dataset, resulting in a lower accuracy. Since the excited state energies are on the order of eV, neglecting snapshots with $|d_{ij}| < \delta$ introduces an error of $\hbar\delta/eV$. For this particular system, $\delta = 1 \times 10^{-3}$ 1/ps ensures sufficient sampling in the area near the conical intersection with an error on the order of $10^{-6}$. 
More numerical details on the choice of this parameter can be found in Sec.~S1.1.1 in the SI.
Then, the dataset is filtered again based on the dihedral angle defined as \chemfig{C-[,0.5]N=[,0.5]N-[,0.5]C}, such that all dihedral angles from 90\degree~ to 180\degree~ have the same weights in the dataset.
The excited-state potential energies,  transition dipoles, and NACRs are then collected.
Additionally, the ground-state potential energies are added to the dataset as well to compute the absorption spectra.
The final dataset includes 24,000 points, which are split into fractions of 7:2:1 for training, validation, and testing sets. Table 1 of a recent paper~\cite{Musaelian2023} shows the performance of nine atomistic machine learning approaches on the MD 17 datasets~\cite{Chmiela2017}, which consist of molecular dynamics trajectories for small molecules, using 1000 configurations for training. The HIP-NN method has itself been tested in this way in prior work to match the PES and found to perform well.~\cite{Lubbers2018JCP} Therefore, the number of snapshots is not a limiting factor in the training.
More details on the dataset preparation process can be found in Section~S1.1 in the Supporting Information (SI).

\begin{figure*}
  \centering
  \includegraphics[width=0.95\textwidth]{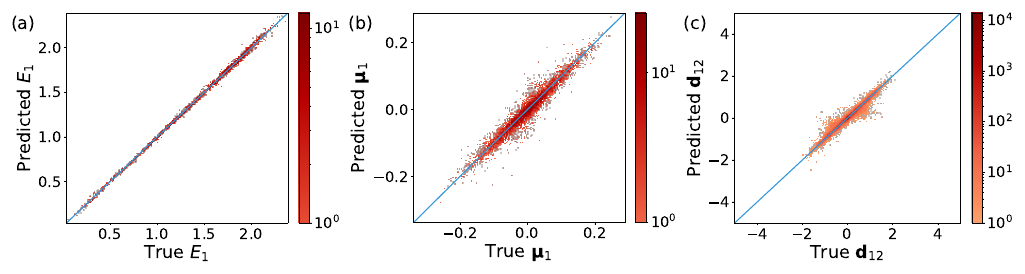}
  \caption{ML model for predicting (a) $S_1$ energies in eV, (b) transition dipoles for $S_0$ to $S_1$ in a.u., and
  (c) NACR between $S_1$ and $S_2$ in 1/Bohr, for the azomethane molecule. The geometries are taken from the
  test set. Note that the phase correction is already applied to $\boldsymbol{\mu}_1$ and
  $\boldsymbol{d}_{12}$.}
  \label{fig:predictions}
\end{figure*}

\subsection{Training}
\label{sec:training}

The learnable parameters in the network (weights $\boldsymbol{w}$, and biases $\boldsymbol{b}$, etc.) need to be assigned properly such that the network can produce accurate predictions.
To obtain these parameters, we minimize a loss function that characterizes the total error of the network.
In the current implementation, we construct the total loss function as a linear combination of all targets we are interested in (molecular energies $E_i$, transition dipoles $\boldsymbol{\mu}_i$, or non-adiabatic coupling vectors $\boldsymbol{d}_{ij}$),
\begin{align}
  \label{eq:total_loss}
  \mathcal{L} = \sum_{y \in \{E, \boldsymbol{\mu}, \boldsymbol{d}\}} \!\!\!\! \lambda_y \mathcal{L}_y + \mathcal{L}_{L2}
\end{align}
where $\lambda_y$ is the weight for the corresponding loss function. As the magnitudes of the loss
functions of different targets vary, this weight is used to balance all targets involved, such that
the predictions are equally accurate for all quantities considered.

The loss function for each target is a combination of the mean absolute errors (MAE) and root mean
square errors (RMSE) of the predicted target $\tilde{y}$ compared to the true value $y$,
\begin{align}
  \label{eq:loss}
  \mathcal{L}_y = \mathrm{MAE}(\tilde{y},y) + \frac{1}{\sqrt{N}} \mathrm{RMSE}(\tilde{y},y),
\end{align}
where $N$ is 1 for scalar quantities (like energies) or the dimensionality for vector
quantities. For example, $N=3$ for dipoles. Note that for the transition dipoles and NACR, the phase-less version of MAE and RMSE is used. An $L2$ term is used to regularize the training,
\begin{align}
\mathcal{L}_{L2} = \lambda_{L2} \lVert \boldsymbol{w} \rVert^2_2,
\end{align}
where $\lVert \boldsymbol{w} \rVert^2_2$ is the square of the L2 norm of all of the weights in the network, $\boldsymbol{w}$.
$\lambda_\mathrm{L2}$ is a user-defined parameter that adjusts the magnitude of the penalty.
The hyper-parameters (model and training parameters that cannot be learned from the dataset) are then obtained from the hyper-parameter search, which is documented in the SI.

The model presented here consists of three interaction layers, each followed by three on-site layers.
Each layer contains 30 neurons. 28 sensitivity functions are used with a lower cutoff distance of 0.77~\AA~ and a cutoff distance of 4.69~\AA. 
The loss function described above is minimized with the ADAM algorithm~\cite{Kingma2017} implemented in the PyTorch package.~\cite{pytorch}
The regularization parameter $\lambda_\mathrm{L2}$ is set to $2 \times 10^{-5}$.
The initial batch size is 32, with a maximum batch size set to 2048. The initial learning rate is $1 \times 10^{-3}$. The model is validated on the validation set at the end of every training epoch.
If the validation error has not been improved for a number of epochs, which is chosen to be 60 here, the batch size will be increased by a factor of 2 until the maximal batch size is reached, and after that, the learning rate will be decreased by a factor of 2.~\cite{Smith2018Arxiv}
The training process will stop earlier if there is no decrease in the validation loss in 120 epochs.
In total, 2,071 epochs are run in the training process.
The training details can be found in Section~S1.2 in the SI.

\subsection{Predictions}\label{sec:pred}

Figure~\ref{fig:predictions} shows the predicted potential energies, transition dipoles, and NACR with input geometries taken from the test set.
Figure~\ref{fig:predictions}a shows the potential energies for $S_1$ with an MAE of 0.026 eV.
Note that the minimum of S$_0$ has been subtracted from all energies in the dataset.
Figure~\ref{fig:predictions}b presents all 3 dimensions of the predicted transition dipoles from $S_0$ to $S_1$ against the NEXMD values with an MAE of 0.012 a.u.
We have applied a phase fix to the predicted dipoles to align the histograms on the main diagonal.
Figure~\ref{fig:predictions}c shows the 2D histogram of the NACR between state $S_1$ and $S_2$ with an MAE of 0.199 1/Bohr.
All 30 dimensions in $\boldsymbol{d}_{12}$ are flattened into a 1D array for plotting purposes.
Due to the complexity of the NACR vector, the predictions' quality is not as good as that for energies or dipoles.
As recent works have pointed out,~\cite{Chen2018JPCL,Chen2023M} when the energy gaps between two states become small, the predicted NACR will have much larger errors. This is indeed an expected behavior in our approach as well. Predicting $\boldsymbol{d}_{ij}(E_j - E_i)$ avoids the singularity issue, but the energy difference $E_j - E_i$ still shows up in the expression nevertheless. When two states are close ($E_j - E_i \to 0$), a small error in energy prediction can cause a huge error to NACR. Such a scenario is also encountered in the explicit electronic structure calculations. A tiny difference in the convergence of energies near the conical interactions will also result in a larger difference in NACRs near conical intersections.  However, the NACR (nearly singularity) is large enough to ensure a high hopping probability when the two states are close. The NACR error near the conical intersections will not statistically translate into NAMD simulation errors. 
Overall, we have achieved accurate predictions for all 3 quantities and all 3 states with one neural network. More comparisons between the predicted results and true results can be found in Section~S2 in the SI.

\subsubsection{Polariton potential energy surface}

Eq.~\ref{eq:jc_N} can be rewritten into the matrix form in the basis of $\{\ket{S_i, 0}, \ket{S_0,
1}\}$,
\begin{align}
  \mathcal{H} =
    \begin{bmatrix}
    \omega_c & g_1 & \dots & g_n \\
    g_1 & \Delta E_1 & \dots & 0 \\ \vdots & \vdots & \ddots & \vdots \\
    g_n & 0 & \dots & \Delta E_n
    \end{bmatrix}.
\end{align}
$\ket{S_i, 0}$ represents the $i^\mathrm{th}$ molecule (for TC) or $i^\mathrm{th}$ state (for JC) is excited, and the photon is in its ground state.
Similarly, $\ket{S_0, 1}$ represents the molecule(s) is in the ground state, and the photon is excited. 
By diagonalizing the matrix, we can obtain the polariton energies. In the case of one two-level molecule coupled to the cavity mode (Eq.~\ref{eq:jc}), the Hamiltonian matrix becomes a $2\times 2$ matrix, such that the energies are immediately available.

\begin{figure}[!htp]
    \centering
    \includegraphics[width=0.45\textwidth]{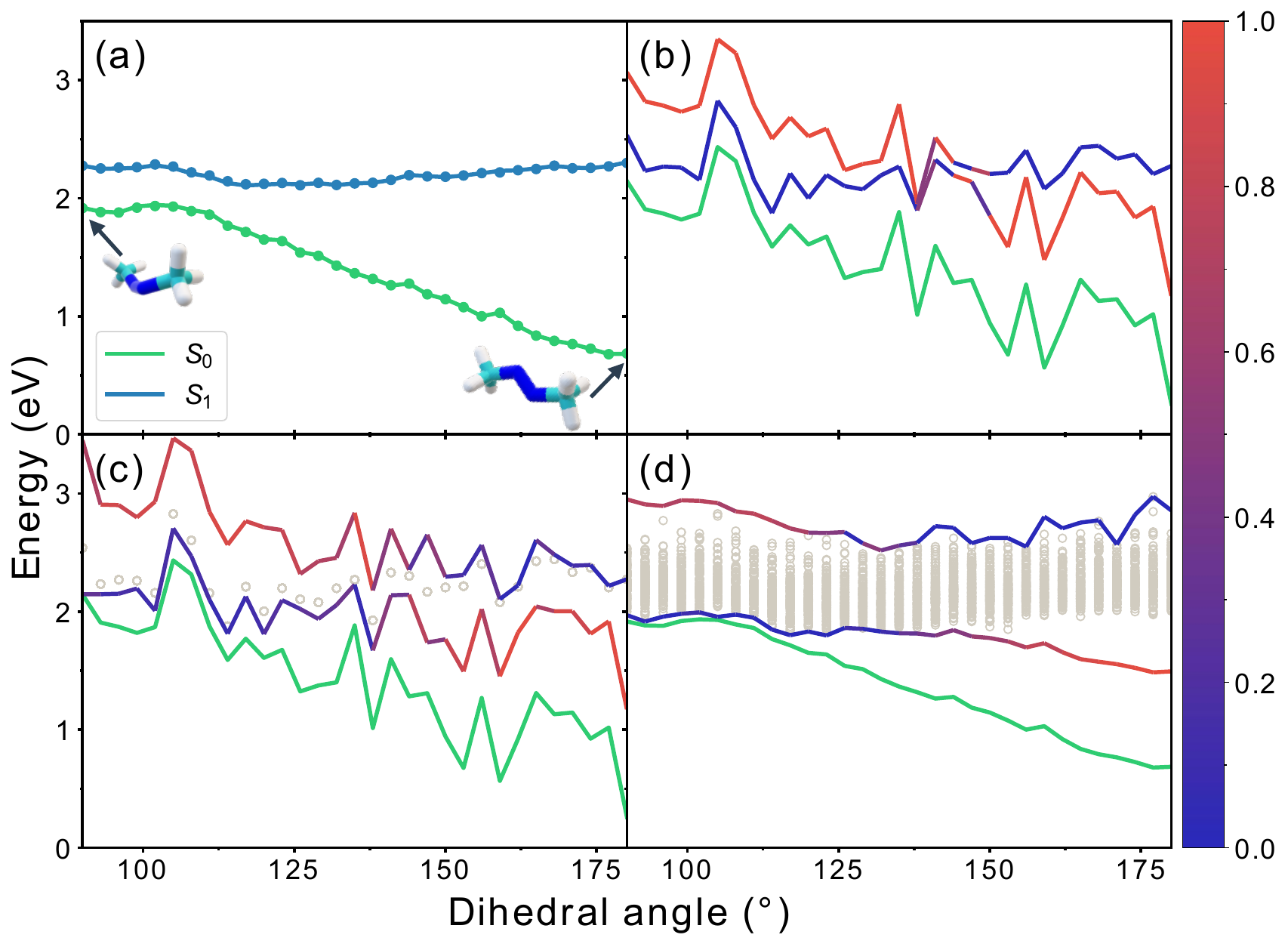}
    \caption{
    Potential energy surface (PES) scan with respect to the dihedral angle defined as \protect\chemfig{C-[,0.5]N=[,0.5]N-[,0.5]C} from 90\degree~ to 180\degree~.
    180\degree~ and 0\degree~ correspond to the trans and cis configurations, respectively.
    Note that the S$_0$ and S$_1$ conical intersection is at 90\degree, at which point the ultrafast part of photo induced dynamics that is of interest to this work is completed.
    (a) The PES of the bare molecule averaged from 100 configurations at each point. The dotted lines are the PES calculated from NEXMD, and the solid lines are from NN predictions. (b-d) The ground state (solid green line) and upper and lower polaritons (colored lines) at various conditions.
    The color bar represents the contribution of the photonic excitations in the polariton state, with 0 being pure electronic excited state ($\ket{S_i, 0}$) and 1 being pure photonic ($\ket{S_0, 1}$).
    The coupling strength parameter is chosen as $g = 0.25$, and photon frequency is chosen to be the energy gap between S$_1$ and S$_0$ at 135\degree. (b) 1 molecule coupled to 1 photon mode case.
    (c) and (d) plot PES of 100 molecules with identical or different configurations coupled to 1 cavity mode, respectively. 
    The molecules are assumed to be aligned to the direction of the electrical field subject to a disorder. The angles between the transition dipoles and the field reflecting disorder effects are modeled by a Gaussian distribution with an FWHM of 60\degree. The grey dots represent all polariton states other than the upper and lower polaritons. Note that for (c), all these states are nearly degenerate.} 
    \label{fig:pes}
\end{figure}

Figure~\ref{fig:pes}a shows the average PES of the ground (S$_0$) and first excited (S$_1$) states averaged from 100 configurations as the azomethane molecule undergoes a trans-cis isomerization.
The details on the geometries used in this figure are provided in the SI.
The predicted results (solid lines) agree very well with the energies calculated from NEXMD (dots).
Note that the S$_0$ and S$_1$ conical intersection is at 90\degree, at which point the ultrafast part of photo induced dynamics that is of interest to this work is completed, such that only trans configurations are included in the plots. 
The subsequent relaxation from the excited state S1 to the ground state S0 occurs through a conical intersection where the CIS method fails to provide correct topology.~\cite{Herbert2022} Assessing such dynamics may be problematic for the standard NAMD simulations with CIS PES. Therefore, we apply a simple model based on energy gaps to qualitatively describe when $S_1\rightarrow S_0$ occurs.~\cite{Zhang2019}
The PES clearly shows that the photoisomerization between the trans and cis isomers can be readily achieved by exciting the molecule to S$_1$.~\cite{Schultz2003}
Figure~\ref{fig:pes}b-d present polariton PES in various cases with the photon frequency chosen as the energy gap between S$_0$ and S$_1$ at 135\degree~ (0.91 eV for b-c  and 0.81 eV for d) and light-matter coupling parameter $g = 0.25$. 
In Figure~\ref{fig:pes}b, only one geometry is used at each dihedral angle.
The two color-mapped plots correspond to the lower and upper polaritons.
The color map represents the photonic contributions in the polariton states, with 0 and 1 being the pure electronic ($\ket{S_i, 0}$) and pure photonic ($\ket{S_0, 1}$) excitations, respectively.
An avoided crossing between the two polaritons is clearly visible at 135\degree.

Figure~\ref{fig:pes}c shows the case where 100 molecules with an identical geometry are collectively coupled to the cavity mode. 
When coupling to the cavity mode, due to the strong interaction between the transition dipoles and the electric field in the cavity, the molecules will have the tendency to align along the direction of the field.
Namely, their transition dipoles will align with the field.
However, realistically, the molecules are not always perfectly aligned, and the system will be subjected to disorder effects.
The angular distribution between the transition dipoles and the field in the cavity should be centered around 0, with the weight gradually vanishing towards the larger angles.
As a result, a Gaussian function is a reasonable choice to describe this distribution qualitatively.
This random alignment is chosen to demonstrate the effect of possible disorders in the polariton eigenstates.
Besides, in a realistic system, the disorders may originate from the dipole misalignment, thermal fluctuations of molecular adiabatic states and their dipoles, and heterogeneous electromagnetic fields in cavities (especially in plasmonic cavities).~\cite{Timmer2023}
Here, we utilize a random distribution of dipole alignment to qualitatively demonstrate the effects of possible disorders.
More detailed and quantitative studies of specific disorders will be discussed in a follow-up paper.
Here, a Gaussian function centered around 0 with a full width at half maximum (FWHM) of 60\degree~ is used. 
The grey dots represent eigenvalues other than the lower and upper polaritons. Even though the random angles can slightly modify the light-matter coupling strength, these eigenvalues are still nearly degenerate and remain unchanged compared to the original S$_1$ PES.
Note that these states will be completely dark on the spectrum, hence called the dark states.
Due to the larger effective coupling strength compared to the 1 molecule case in Figure~\ref{fig:pes}a, the Rabi splitting is much larger.
As a result, the energy difference between upper and lower polaritons is more visible across all dihedral angles.
The plots in Figure~\ref{fig:pes}d are the eigenvalues calculated from 100 molecules with different configurations coupled to the cavity mode, with random angles with an FWHM of 60\degree.
As all molecules have very different coupling strengths, the states represented by the grey dots are not degenerate anymore.
At the avoided crossing at 135\degree, both upper and lower polaritons have a 50\% mixing of photonic and electronic excitations.
Both states become either pure photonic or electronic, moving away from the midpoint.

\subsection{Absorption spectrum}

The absorption spectrum for a given configuration with Lorentzian broadening can be calculated via 
\begin{align}
  A_{i}^{k}\left(\epsilon\right) =
    \frac{4}{3\pi\Gamma} \Omega_i \lVert \boldsymbol{\mu}_i \rVert^2
    \frac{1}{1+4\left(\frac{\epsilon-\Omega_i}{\Gamma}\right)^{2}},
\end{align}
where $A_i^k$ is the absorption for the $i^\mathrm{th}$ state and the $k^\mathrm{th}$ configuration, $\Omega_i$
is the excitation energy from the ground state $S_0$ to the $i^\mathrm{th}$ excited state $S_i$, and
$\Gamma$ is the Lorentzian broadening parameter. The total absorption spectrum of the molecule can
be obtained by averaging an ensemble of configurations,
\begin{align}
  A_i\left(\epsilon\right) = \frac{1}{N} \sum_{k=1}^N A_i^k \left(\epsilon\right).
\end{align}

\begin{figure}
    \centering
    \includegraphics[width=0.45\textwidth]{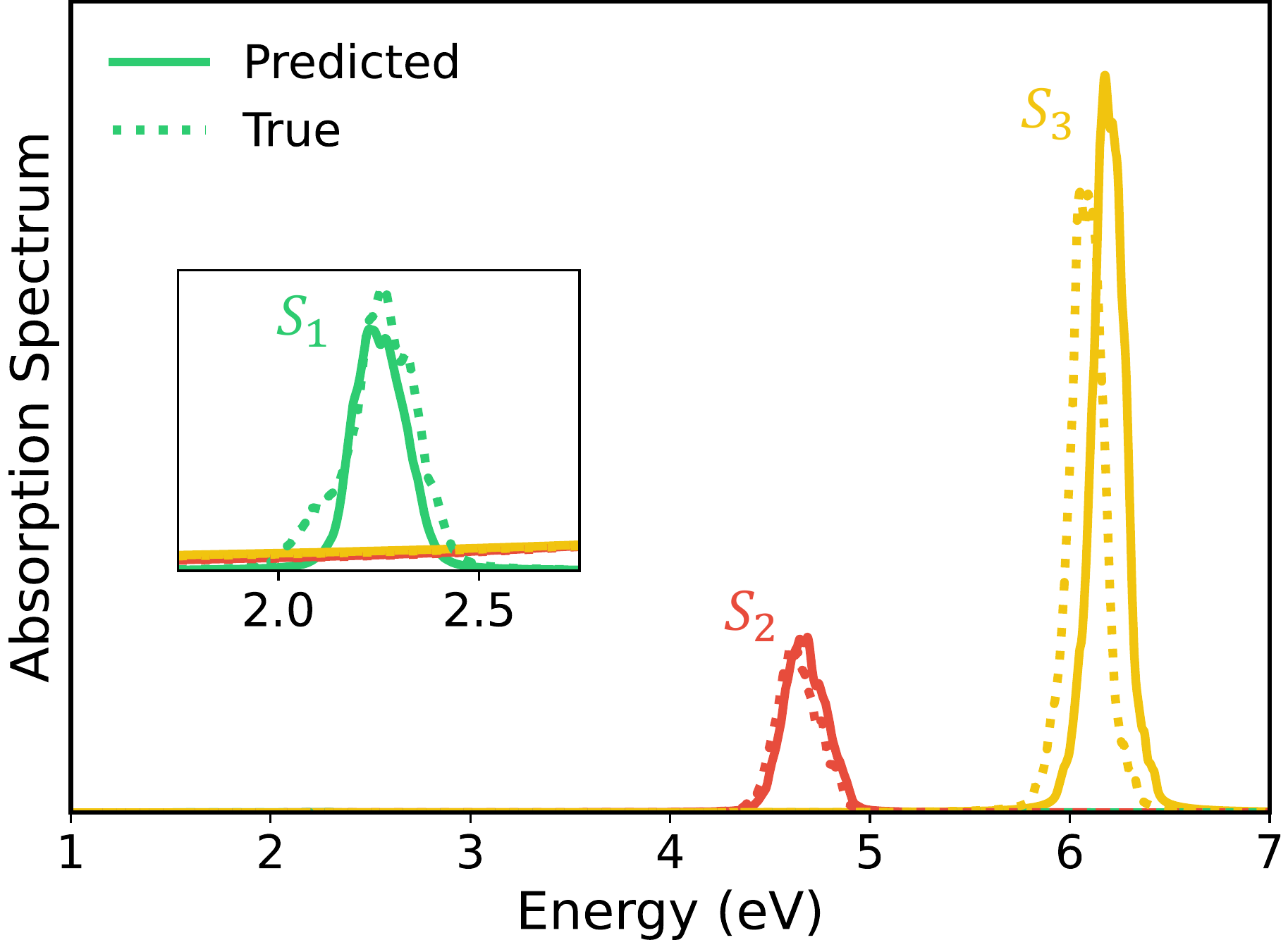}
    \caption{Absorption spectrum of the first three excited states of the azomethane molecule subject to thermal fluctuations. The dotted lines represent the spectrum calculated from NEXMD simulations, and the solid lines are calculated from NN predictions.}
    \label{fig:absorption}
\end{figure}

The configurations used to calculate the spectra are sampled from a ground-state molecular dynamics simulation at ambient conditions with the details provided in Section S1.1 in the SI. 
To reproduce the spectrum at 300 K, the broadening factor is chosen as $\Gamma = 0.02585$ eV, corresponding to the thermal energy at 300 K. Figure~\ref{fig:absorption} shows the absorption spectrum of all 3 excited states considered in this model. Note that $\boldsymbol{\mu}_1$ is significantly smaller than the $\boldsymbol{\mu}_2$ and $\boldsymbol{\mu}_3$. As a result, the absorption of S$_1$ is much smaller with the details shown in the zoom-in part. We use 1,000 configurations close to the optimized ground-state configuration as the input data, which does not exist in our training dataset. As our model can accurately predict the excitation energies and transition dipoles, both the energies corresponding to the absorption peak and the absorption magnitudes are close to the values calculated from NEXMD simulations.

\begin{figure}
    \centering
    \includegraphics[width=0.45\textwidth]{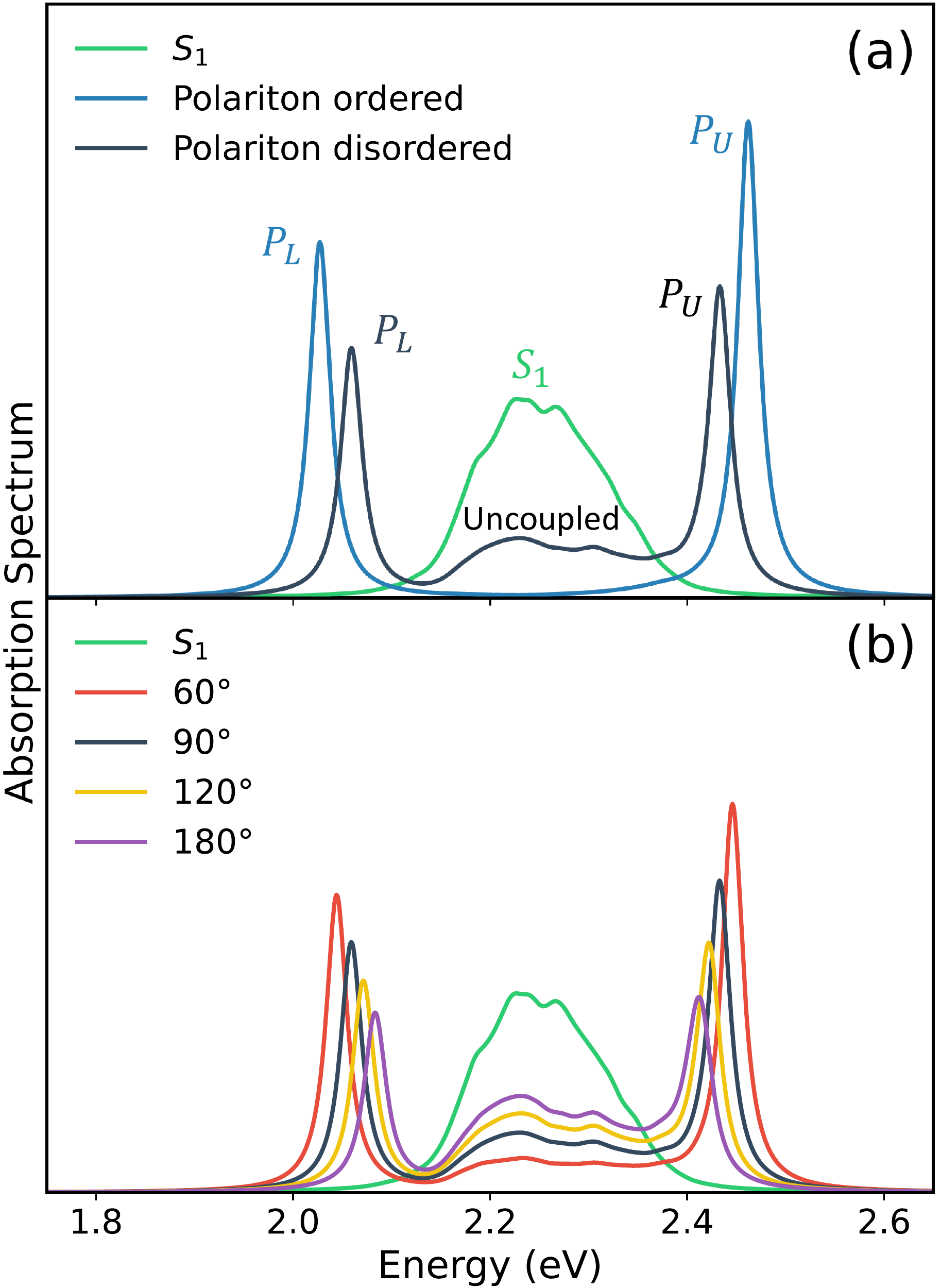}
    \caption{
    The absorption spectra of the first excited state (S$_1$) (green) and polariton states calculated at various levels of disorder.
    1,000 molecules are collectively coupled to the cavity mode, and the light-matter coupling parameter is set to $g = 0.15$.
    The geometries of these molecules are sampled from a ground-state molecular dynamics simulation with details provided in Section~S1.1 in the SI.
    (a) Comparison between completely aligned transition dipoles (blue) and dipoles with a dispersion disorder (black). The disorder is modeled with a Gaussian function centered around 0 with an FWHM of 90\degree. When the angle between the transition dipole and the electric field is close to 90\degree, the light-matter coupling strength will be close to 0, such that uncoupled molecules are showing up in the spectrum. (b) The effects of increasing disorder on the polariton spectra. As the FWHM increases from 60\degree~ to 180\degree~, the Rabi splitting becomes smaller, and more uncoupled molecules show up.}
    \label{fig:polariton_spectrum}
\end{figure}

Figure~\ref{fig:polariton_spectrum}a shows the bare-molecule absorption spectrum of S$_1$ and the corresponding polariton spectra in two scenarios, when all transition dipoles $\boldsymbol{\mu}_1$ are ordered (blue), i.e., the transition dipoles are aligned with the electric field, or the dipoles are disordered (black).
The disorder is modeled by a Gaussian function centered around 0 with an FWHM of 90\degree.
When the transition dipole is perpendicular to the electric field in the cavity, the coupling strength becomes 0 (Eq.~\ref{eq:gc_dot}).
The disorder has two impacts on the polariton spectrum. First, the random angles will reduce the overall effective coupling strength for all molecules, resulting in a smaller Rabi splitting and lower absorption peak for both polaritons on the spectrum.
Second, without light-matter coupling, these molecules' transition dipoles and excitation energies will remain unchanged.
Therefore, their absorption spectrum is exactly the same as the bare molecule case, and the third peak labeled as "uncoupled" in Figure~\ref{fig:polariton_spectrum}a emerges.
We must emphasize that this effect intrinsically arises from the scenario where many molecules couple to the cavity mode, and there is a disorder in the orientation of these molecules. As a result, this, by no means, can be understood from simulations that only include one molecule or a simple model system where the molecules are assumed to be uniform and ordered.
As we have discussed, even with semi-empirical Hamiltonians and MQC methods, simulating many realistic molecules coupled to a cavity mode can still be computationally expensive. Thus, accelerating such simulations with ML models is a rational choice.
Figure~\ref{fig:polariton_spectrum}b shows that as the magnitude of the disorder increases, the Rabi splitting decreases, and more uncoupled molecules show up on the spectra.

\section{Conclusion}
To summarize, we have developed a comprehensive protocol to predict excited-state properties, including energies, transition dipoles, and non-adiabatic coupling vectors, for a given molecule, using the Hierarchically Interacting Particle Neural Network, encapsulated within an open-source Python package \texttt{hippynn}.
In the study of light-matter interactions, particularly in the realm of polariton chemistry of many molecules, our model offers significant advantages. Polariton states are hybridized states formed between photons and many molecules within an optical cavity. Each molecule within this system might assume various configurations while maintaining identical chemical compositions.

Our protocol, having been trained on a single molecule in an ample configuration space, enables accurate predictions of the properties of these molecules across different configurations. Therefore, it circumvents the computationally intensive task of explicitly computing the energies, transition dipoles, and NACRs of many chemically identical molecules, significantly reducing computational costs for future non-adiabatic dynamics simulations for exciton-polaritons compared to the reference quantum-mechanical simulations. We have demonstrated the efficacy of this approach using a prototype molecule, azomethane. The model accurately reproduces spectra for individual molecules and molecular polaritons. Moreover, by integrating our trained model with existing non-adiabatic molecular dynamics packages, it will be feasible to conduct simulations of excited-state polariton chemistry where many molecules couple to the cavity modes. This bypasses the need for resource-intensive electronic structure calculations, potentially speeding up simulations of polaritonic systems by several orders of magnitude. As a result, this advancement could lead to a more in-depth understanding of polariton chemistry and potentially unveil new aspects of light-matter interactions.

\begin{acknowledgement}
The authors thank Sakib Matin, Benjamin Nebgen, and Philipp Marquetand for valuable discussions and feedback. We acknowledge support from the US DOE, Office of Science, Basic Energy Sciences, Chemical Sciences, Geosciences, and Biosciences Division under Triad National Security, LLC (``Triad'') contract Grant 89233218CNA000001 (FWP: LANLE3F2). 
The research is performed in part at the Center for Integrated Nanotechnologies (CINT), a U.S. Department of Energy, Office of Science user facility at LANL. 
This research used resources provided by the LANL Institutional Computing (IC) Program, which is supported by the U.S. Department of Energy National Nuclear Security Administration under Contract No. 89233218CNA000001, and the Darwin testbed which is funded by the Computational Systems and Software Environments subprogram of LANL's Advanced Simulation and Computing program (NNSA/DOE).
LANL is operated by Triad National Security, LLC, for the National Nuclear Security Administration of the U.S. Department of Energy (Contract No. 89233218CNA000001). 
\end{acknowledgement}

\begin{suppinfo}
More details about the neural network model, data generation, training procedure, and additional supporting figures are available in the Supporting Information. An example training script, \texttt{excited\_states\_azomethane.py}, that trains a model to the dataset, is available as part of the {\it hippynn} open-source code at \url{https://github.com/lanl/hippynn}. The dataset for this work is available at \url{https://doi.org/10.5281/zenodo.7076420}.
\end{suppinfo}


%

\bibliography{ref}

\makeatletter\@input{suppinfo_aux.tex}\makeatother
\end{document}


\section{Neural network model details}
\label{sec:training_details}

\subsection{Data generation}
\label{sec:data}

The training data is generated with the NEXMD package~\cite{Malone2020,Nelson2020,Freixas2023} following the common non-adiabatic molecular dynamics (NAMD) procedure. The electronic structure is calculated at the configuration interaction singles (CIS) level with AM1 semi-empirical Hamiltonian.~\cite{Dewar1985} 
Trajectory surface hopping (TSH) is used for the non-adiabatic dynamics.~\cite{Tully2012}
We first run a 1,000,000-step ground-state Born-Oppenheimer molecular dynamics simulations (BOMD) with a time step of 0.1 fs at 300 K. A Langevin thermostat is used to maintain the temperature. We then collect the configurations and velocities every 10 fs (100 steps), resulting in 10,000 total configurations. Among them, 9,000 are used as the initial conditions for subsequent NAMD simulations, and the remaining 1,000 configurations are saved for the absorption spectrum calculations in Figures~5 and 6 in the main text.

All trajectories are initially excited to the first excited state ($S_1$), with the first 5 excited states included in the simulations. Trajectory surface hopping~\cite{Tully1971} was chosen to run the trajectories. The time step for the classical nuclei is 0.1 fs, and within each classical time step, 4 quantum steps are used to propagate the electronic degrees of freedom (DOFs). Each trajectory will run 10,000 steps or stop early whenever the molecule hops back to the ground state. The geometries, energies (including the ground state and all excited states), transition dipoles, and non-adiabatic coupling vectors (NACRs) are collected into the dataset. The snapshot is added to the dataset every 10 steps or whenever there is a strong coupling between any pairs of excited states. The strong coupling is determined by $ |d_{ij}| > 1 \times 10^{-3}$ 1/ps, where $|d_{ij}|$ is the absolute value of the non-adiabatic coupling term  (NACT). We then filter the resulting dataset based on the dihedral angles defined by \chemfig{C-[,0.5]N=[,0.5]N-[,0.5]C}, such that all dihedral angles from 90\degree~ to 180\degree~ have approximately the same occurrence in the dataset. The final dataset has about 96,000 snapshots.

With the collected dataset, 1/4 of the snapshots were used in the training process (details in the next section), and 3/4 were saved for the PSE scan.
We scanned from 90\degree~ to 180\degree~ with an increment of 3\degree, resulting in 31 different dihedral angles.
At each angle, we select 100 snapshots whose dihedral angle is closest to the designated angle.
These snapshots are used to plot Figure~4a and d in the main text.
Figure~4b-c in the main text only use the geometry with the closest dihedral angle.

\subsubsection{Impact of the threshold $\delta$}\label{sec:delta}

\begin{figure}
  \centering
  \includegraphics[width=0.95\textwidth]{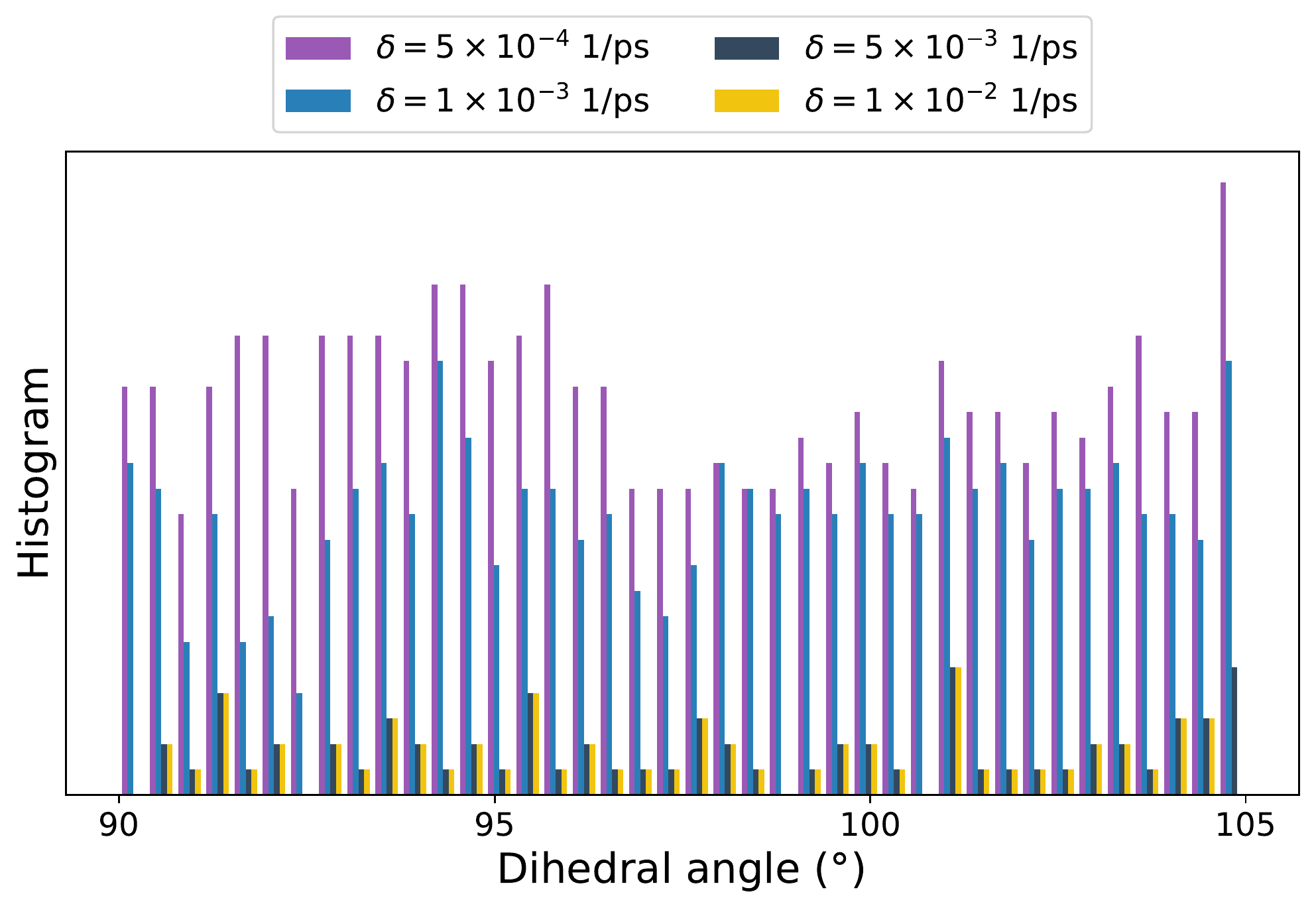}
  \caption{Un-normalized histograms of the dihedral angle defined by \protect\chemfig{C-[,0.5]N=[,0.5]N-[,0.5]C} near the conical intersection (from 90\degree~ to 105\degree) with 4 different threshold before filtering. $\delta=1\times 10^{-3}$ 1/ps gives better and more balanced coverage over this region.}
  \label{fig:delta}
\end{figure}

As the trajectories only stay near the conical intersection for a short period of time, it is very difficult to ensure enough sampling.
Smaller threshold $\delta$ results in more configurations being sampled across the board, including the conical intersection area.
As Fig.~\ref{fig:delta} shows, a smaller $\delta$ value gives better and more balanced coverage close to 90\degree, even before we perform the filtering.
$\delta=5\times 10^{-4}$ performs very similarly to $\delta=1\times 10^{-3}$, except about 25\% increase in the total number of samples, ultimately becoming a user preference on which one to choose.
It should also be noted that $\delta$ going further small will cause a quick increase in the number of snapshots near the equilibrium but barely any gain near the conical intersection.

\subsection{Training details}

There are two patience controller parameters in {\it hippynn}. At the end of every training epoch,
the model is validated on the validation set. If the validation error has not been improved for a
number of epochs, the batch size will be increased by a factor of 2 until the maximal batch size is
reached, and after that, the learning rate will be decreased by a factor of 2.~\cite{Smith2018Arxiv}
This parameter is called batch size raising patience. The other parameter is called early stopping
patience. The training process will stop early if there is no decrease in the user-selected
validation metric in the number of epochs defined by this parameter. In the whole training process,
the total validation loss is chosen as the metric for early stopping.

Only 1/4 of the snapshots (24,000) are used in the training process with a training, validation, and
test splitting of 7:2:1. Ax package~\cite{ax} is used to perform hyper-parameter searching. Ray
package ~\cite{ray} is used in conjunction with Ax to parallelize the search on multiple GPUs, hence
improving search efficiency. The whole search was split into two steps. First, we search for the
optimal cutoff distances. 40 searches were performed in total in this step. The batch size raising
patience and early stopping patience are 40 and 100 epochs, respectively. Then, we optimize the
weights for all 3 quantities (energy, dipole, and NACR) in the loss function. The patience for
raising batch size is also increased to 60 epochs, such that more epochs can be allowed in the
training. 20 searches were performed in this step.

The final network presented in this work has three interaction layers, and each is followed by three
atomic layers. Each layer contains 30 neurons. 28 sensitivity functions are used with a lower cutoff
distance (midpoint of first sensitivity function) of 0.77~\AA, upper cutoff (midpoint of last
sensitivity function) of 3.41~\AA, and a cutoff distance (cutoff for all sensitivity functions) of
4.69~\AA. The loss function defined in Eq.~19 in the main text is minimized with
the ADAM algorithm~\cite{Kingma2017} implemented in the PyTorch package.~\cite{pytorch}  The initial
learning rate is chosen to be $1 \times 10^{-3}$, and all other parameters for ADAM are the default
settings of PyTorch. The regularization parameter $\lambda_{L2}$ is set to $2 \times 10^{-5}$, and
the weights ($\lambda_y$) for energy, dipole, and NACR targets are 1:4:2. The initial batch size is
chosen to be 32, with a maximum batch size set to 2048. The
model is validated on the validation set at the end of every training epoch. With the batch size raising patience and early stopping
patience set to 60 and 120 epochs, in total, 2071 epochs are performed in the training process.

\subsection{Metrics}

The details of all training metrics are listed below in Table~\ref{tbl:metrics}. Good accuracy is
achieved for all quantities and states included in the training.

\begin{table}[!htp]
  \begin{tabular}{|l|r|r|r|}
    \hline
                & \multicolumn{1}{|c|}{Training} & \multicolumn{1}{|c|}{Validation} & \multicolumn{1}{|c|}{Test} \\ \hline
    \multicolumn{1}{|c|}{Energy RMSE} &  0.02754 & 0.028264 & 0.028092 \\ \hline
    \multicolumn{1}{|c|}{Energy MAE}  &  0.01981 & 0.020353 & 0.020388 \\ \hline
    \multicolumn{1}{|c|}{Energy loss} &  0.04735 & 0.048618 &  0.04848 \\ \hline
    \multicolumn{1}{|c|}{Dipole RMSE} & 0.033204 & 0.036814 & 0.035535 \\ \hline
    \multicolumn{1}{|c|}{Dipole MAE}  & 0.017378 & 0.018479 & 0.018319 \\ \hline
    \multicolumn{1}{|c|}{Dipole loss} & 0.036549 & 0.039734 & 0.038835 \\ \hline
    \multicolumn{1}{|c|}{NACR RMSE}   &  0.22336 &   0.2262 &  0.22761 \\ \hline
    \multicolumn{1}{|c|}{NACR MAE}    &    0.104 &  0.10514 &  0.10632 \\ \hline
    \multicolumn{1}{|c|}{NACR loss}   &  0.14478 &  0.14644 &  0.14788 \\ \hline
    \multicolumn{1}{|c|}{L$_2$ loss}  &   1184.6 &   1184.6 &   1184.6 \\ \hline
    \multicolumn{1}{|c|}{Total loss}  &  0.50679 &  0.52413 &  0.52326 \\ \hline
  \end{tabular}
  \caption{Final training metrics. The units for energy, transition dipoles, and NACR are eV, a.u.,
  1/Bohr, respectively. Each quantity contains all states considered in the training process. The
  loss for each quantity is a weighted sum of the RMSE and MAE defined in Eq.~20 in
  the main text.}
  \label{tbl:metrics}
\end{table}

\section{Training results}\label{sec:histograms}

All histograms show the comparison between the predictions of the test set and the true results
obtained from NEXMD simulations. In each figure, all panels share the same color map indicated by
the color bar on the right.

\begin{figure}
  \centering
  \includegraphics[width=0.95\textwidth]{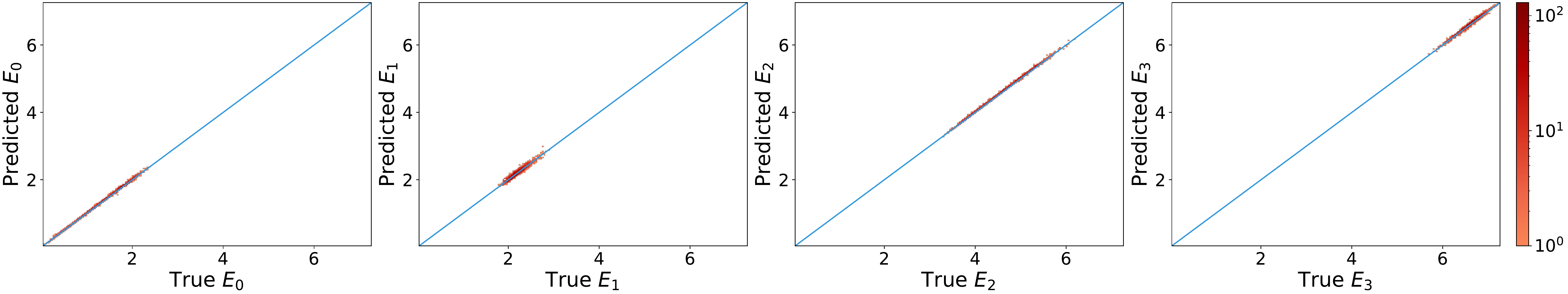}
  \caption{Energy predictions of the ground state and first 3 excited states on the test set. The energy unit is eV.}
  \label{fig:E_predictions}
\end{figure}

\begin{figure}[!htp]
  \centering
  \includegraphics[width=0.73\textwidth]{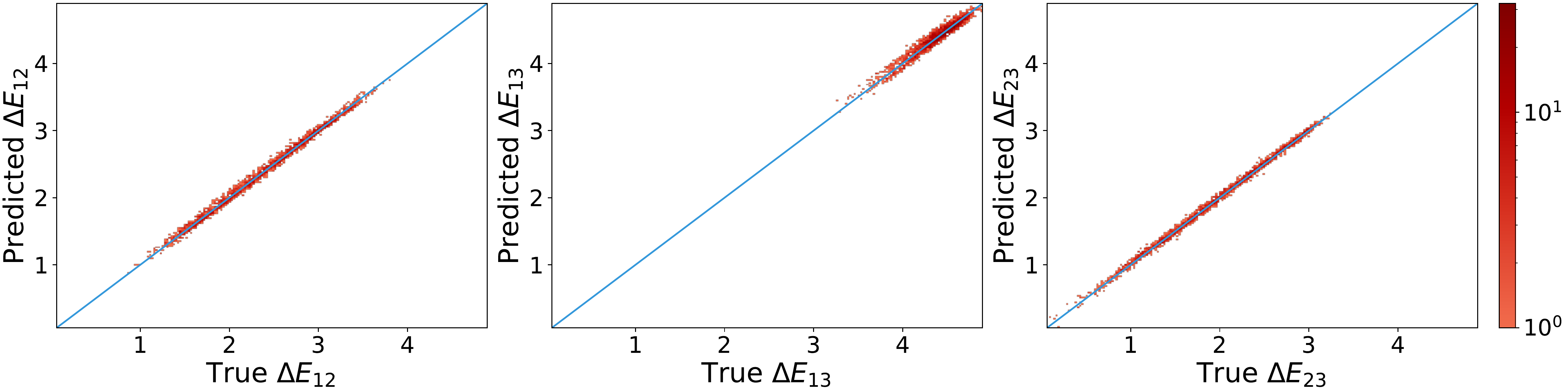}
  \caption{Predicted energy difference between adjacent states compared to the true values in eV.}
  \label{fig:dE_predictions}
\end{figure}

\begin{figure}
  \centering
  \includegraphics[width=0.73\textwidth]{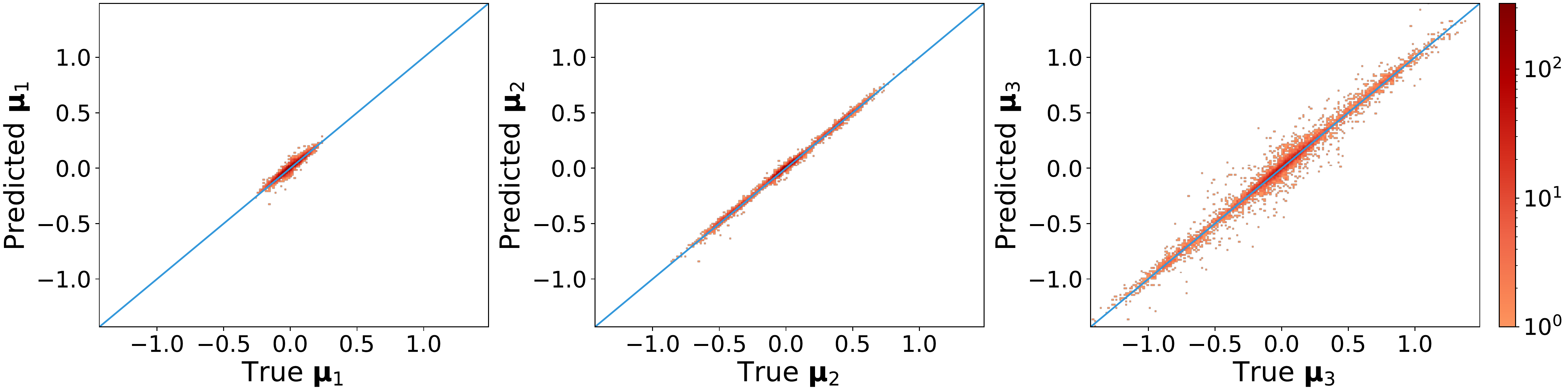}
  \caption{Predicted transition dipole from the ground state to the first 3 excited states on the test set. The dipole unit is a.u.}
  \label{fig:D_predictions}
\end{figure}

\begin{figure}
  \centering
  \includegraphics[width=0.73\textwidth]{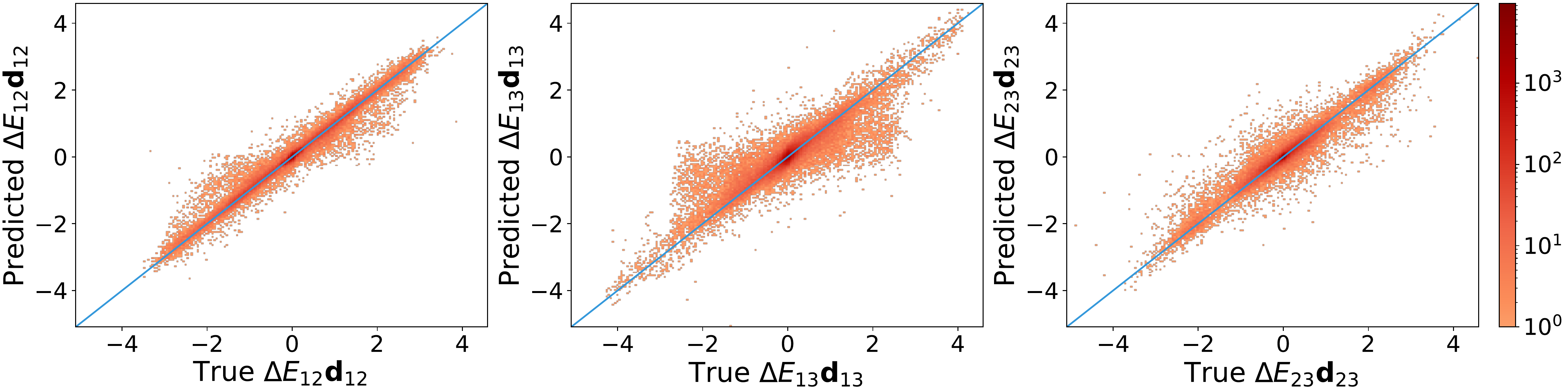}
  \caption{Predicted NACR between all pairs of excited states multiplied by the corresponding energy gaps. The unit is eV/Bohr.}
  \label{fig:NACRdE_predictions}
\end{figure}

\begin{figure}
  \centering
  \includegraphics[width=0.73\textwidth]{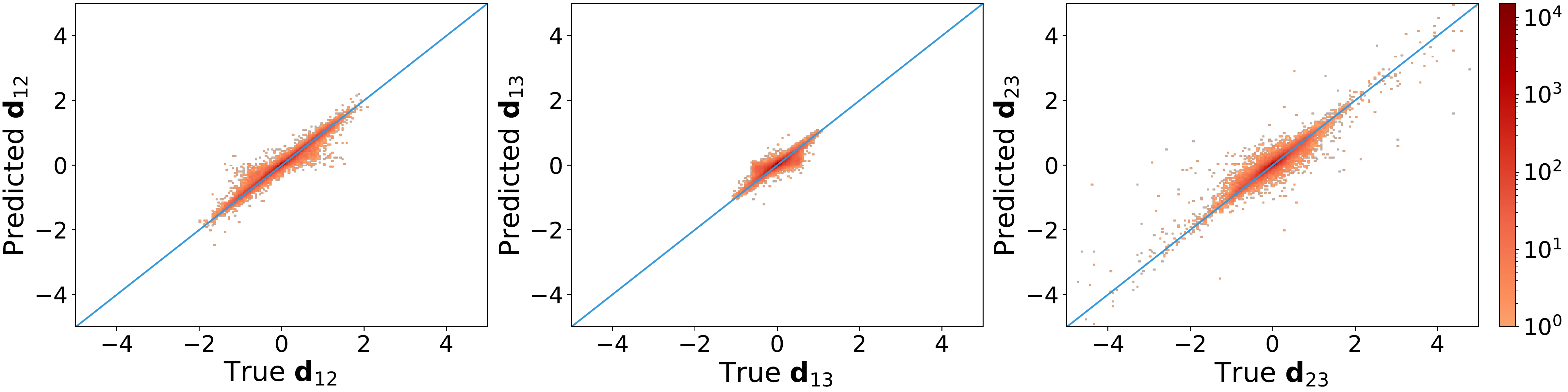}
  \caption{Predicted NACR between all pairs of excited states. The unit is 1/Bohr.}
  \label{fig:NACR_predictions}
\end{figure}

\section{Absorption spectra}

\begin{figure}
  \centering
  \includegraphics[width=0.5\textwidth]{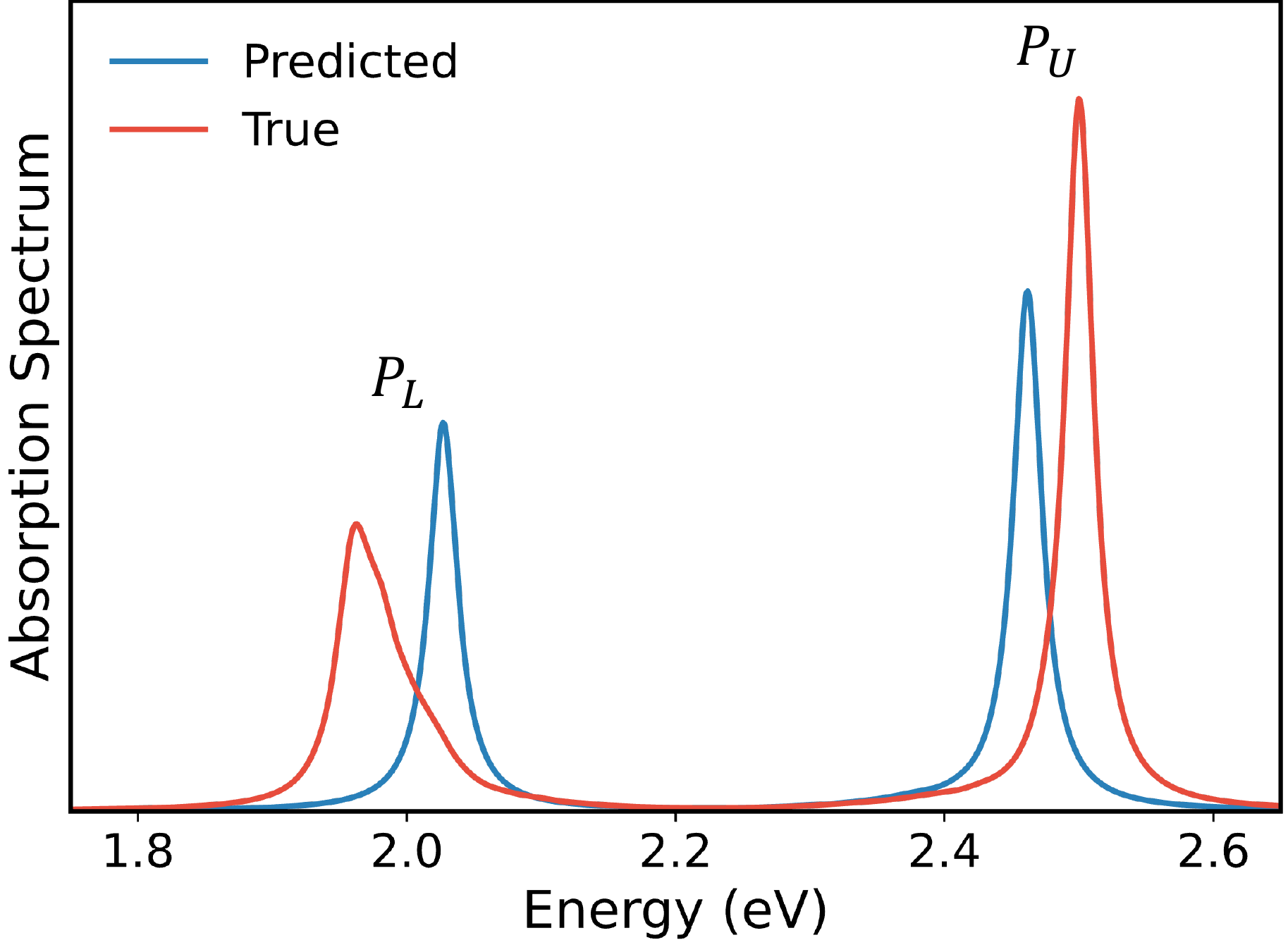}
  \caption{Predicted (blue) and true (red) absorption spectra of the polaritons. 1,000 molecules are coupled to the cavity mode.
  The predicted peak of the lower polariton is narrower, and the predicted Rabi splitting is also smaller.}
  \label{fig:polariton_pred_true}
\end{figure}
%
%

\bibliography{ref}
\makeatletter\@input{main_aux.tex}\makeatother

%% file: command.tex
\usepackage{bm}
\usepackage{braket}
\usepackage{capt-of}
\usepackage{chemfig}
\usepackage{color}
\usepackage{dcolumn}
\usepackage{float}
\usepackage[T1]{fontenc} 
\usepackage{gensymb}
\usepackage{graphicx}
\usepackage{mathrsfs}
\usepackage{tikz}
\usepackage[normalem]{ulem}

\graphicspath{{figures/}}

\definecolor{NNGreen}{RGB}{38, 222, 129}
\definecolor{NNBlue}{RGB}{55, 66, 250}
\definecolor{NNBlack}{RGB}{47, 53, 66}
\definecolor{NNYellow}{RGB}{241, 196, 15}
\definecolor{NNRed}{RGB}{255, 107, 107}

\usetikzlibrary{arrows.meta,
                calc, chains,
                decorations.pathreplacing,
                calligraphy,
                positioning,
                math}
\makeatletter
\def\squarecorner#1{
    \pgf@x=\the\wd\pgfnodeparttextbox%
    \pgfmathsetlength\pgf@xc{\pgfkeysvalueof{/pgf/inner xsep}}%
    \advance\pgf@x by 2\pgf@xc%
    \pgfmathsetlength\pgf@xb{\pgfkeysvalueof{/pgf/minimum width}}%
    \ifdim\pgf@x<\pgf@xb%
        \pgf@x=\pgf@xb%
    \fi%
    \pgf@y=\ht\pgfnodeparttextbox%
    \advance\pgf@y by\dp\pgfnodeparttextbox%
    \pgfmathsetlength\pgf@yc{\pgfkeysvalueof{/pgf/inner ysep}}%
    \advance\pgf@y by 2\pgf@yc%
    \pgfmathsetlength\pgf@yb{\pgfkeysvalueof{/pgf/minimum height}}%
    \ifdim\pgf@y<\pgf@yb%
        \pgf@y=\pgf@yb%
    \fi%
    \ifdim\pgf@x<\pgf@y%
        \pgf@x=\pgf@y%
    \else
        \pgf@y=\pgf@x%
    \fi
    \pgf@x=#1.5\pgf@x%
    \advance\pgf@x by.5\wd\pgfnodeparttextbox%
    \pgfmathsetlength\pgf@xa{\pgfkeysvalueof{/pgf/outer xsep}}%
    \advance\pgf@x by#1\pgf@xa%
    \pgf@y=#1.5\pgf@y%
    \advance\pgf@y by-.5\dp\pgfnodeparttextbox%
    \advance\pgf@y by.5\ht\pgfnodeparttextbox%
    \pgfmathsetlength\pgf@ya{\pgfkeysvalueof{/pgf/outer ysep}}%
    \advance\pgf@y by#1\pgf@ya%
}
\makeatother

\pgfdeclareshape{square}{
    \savedanchor\northeast{\squarecorner{}}
    \savedanchor\southwest{\squarecorner{-}}

    \foreach \x in {east,west} \foreach \y in {north,mid,base,south} {
        \inheritanchor[from=rectangle]{\y\space\x}
    }
    \foreach \x in {east,west,north,mid,base,south,center,text} {
        \inheritanchor[from=rectangle]{\x}
    }
    \inheritanchorborder[from=rectangle]
    \inheritbackgroundpath[from=rectangle]
}
\usepackage{filecontents}

%% file: hipnn_diagram.tex
\begin{figure}
  \scalebox{0.8}{\begin{tikzpicture}[
    font = \sffamily,
    shorten > = 1pt,
    > = Stealth,
    line width = 1.5pt,
    start chain = going below,
    neuron/.style = {
      square,
      fill=#1,
      minimum size= 18pt,
      inner sep = 1pt,
      rounded corners=2pt,
    },
  ]

    \pgfmathsetmacro\nrows{4};
    \pgfmathsetmacro\nint{2};
    \pgfmathsetmacro\natom{3};
    \pgfmathsetmacro\idots{int(\nrows-1)};

    \foreach \i in {1, ..., \nrows} {
      \ifnum \i = \idots
        \node[neuron=white, on chain, below=2mm] (n-1-\i) {$\vdots$};
      \else
        \ifnum \i = \nrows
          \pgfmathsetmacro\idx{"n"};
        \else
          \pgfmathsetmacro\idx{\i};
        \fi
        \node[neuron=NNBlue, on chain, below=3mm] (n-1-\i) {\color{white}$A_\idx$};
      \fi
    }
    \node[neuron=NNBlack, on chain, below=5mm] (e-1) {\color{white}$\tilde{E}_{k,i}^0$};
    \node[neuron=NNBlack, below=5mm of e-1] (q-1) {\color{white}$\tilde{q}_{k,i}^0$};
    \foreach \i in {1, ..., \nint} {
      \pgfmathsetmacro\blockidx{int((\i - 1) * \natom + 1)};
      \path   let \p1 = ($(n-\blockidx-1.north) - (n-\blockidx-\nrows.south)$),
                  \n\i = {veclen(\y1,\x1)} in
              node (r\i) [
                minimum height=\n\i,
                minimum width=7mm, fill = NNGreen,
                below right=0mm and 6 mm of n-\blockidx-1.north,
                rounded corners=2pt
              ]{\rotatebox{90}{Interaction layer}};
      \foreach \j in {1, ..., \natom} {
        \pgfmathsetmacro\nodeidx{int(\blockidx + \j)};
        \ifnum \j = 1
          \pgfmathsetmacro\parent{"r\i"};
        \else
          \pgfmathsetmacro\parentidx{int(\nodeidx - 1)};
          \pgfmathsetmacro\parent{"n-\parentidx-1"};
        \fi
        \foreach \k in {1, ..., \nrows} {
          \ifnum \k = \idots
            \node[neuron=white, right=3mm of n-1-\k -| \parent.east] (n-\nodeidx-\k) {$\vdots$};
          \else
            \node[neuron=NNRed, right=3mm of n-1-\k -| \parent.east] (n-\nodeidx-\k) {};
            \draw[-] (n-\nodeidx-\k) -- (n-1-\k -| \parent.east) ++ (0.5,0);
          \fi
        }
      }
      \pgfmathsetmacro\parent{int(\i * \natom)};
      \pgfmathsetmacro\nodeidx{int(\i + 1)};
      \node[neuron=NNBlack, right=3mm of e-1 -| n-\parent-\nrows.east] (e-\nodeidx) {\color{white}$\tilde{E}_{k,i}^\i$};
      \node[neuron=NNBlack, below=5mm of e-\nodeidx] (q-\nodeidx) {\color{white}$\tilde{q}_{k,i}^\i$};
    }
    \pgfmathsetmacro\nlayers{int(\nint + 1)}
    \pgfmathsetmacro\out{int(\nint + 2)}
    \node[neuron=NNYellow, right=10mm of e-\nlayers] (e-\out) {\color{black}$\tilde{E}_i$};
    \node[neuron=NNBlack, below=5mm of q-\nlayers] (q-\out) {\color{white}$\tilde{\boldsymbol{q}}_i$};

    \foreach \i in {1, ..., \nrows} {
      \ifnum \i = \idots
      \else
        \draw[-] (n-1-\i) -- (n-1-\i.east -| r1.west) ++ (0.5,0);
      \fi
    }
    \foreach \i in {2, ..., \nint} {
      \pgfmathsetmacro\atom{int((\i - 1) * \natom + 1)}
      \foreach \j in {1, ..., \nrows} {
        \ifnum \j = \idots
        \else
          \draw[-] (n-\atom-\j) -- (n-1-\j.east -| r\i.west) ++ (0.5,0);
        \fi
      }
    }
    \foreach \i in {0, ..., \nint} {
      \pgfmathsetmacro\outidx{int(\i + 1)}
      \pgfmathsetmacro\atomidx{int(\i * \natom + 1)}
      \draw[-] (e-\outidx.north) -- (n-\atomidx-\nrows.south);
      \draw[-] (q-\outidx.north) -- (e-\outidx.south);
      \ifnum \i > 0
        \pgfmathsetmacro\atomidx{int((\i - 1) * \natom + 2)}
        \node[circle, minimum size=12pt, draw=NNBlack, at = (e-1 -| n-\atomidx-1)] (pe-\i) {};
        \node[
          circle, minimum size=8pt, at = (e-1 -| n-\atomidx-1),
          path picture={
        \draw[NNBlack] (path picture bounding box.west)+(1pt,0) -- (path picture bounding box.east); 
        \draw[NNBlack] (path picture bounding box.north)+(0,-1pt) -- (path picture bounding
        box.south); 
          }
        ] () {};
        \draw[-] (e-\outidx) -- (pe-\i);
        \draw[-] (e-\i) -- (pe-\i);
        \node[circle, minimum size=12pt, draw=NNBlack, at = (q-2 -| n-\atomidx-1)] (pq-\i) {};
        \node[
          circle, minimum size=8pt, at = (q-2 -| n-\atomidx-1),
          path picture={
        \draw[NNBlack] (path picture bounding box.west)+(1pt,0) -- (path picture bounding box.east); 
        \draw[NNBlack] (path picture bounding box.north)+(0,-1pt) -- (path picture bounding
        box.south); 
          }
        ] () {};
        \draw[-] (q-\outidx) -- (pq-\i);
        \draw[-] (q-\i) -- (pq-\i);
      \fi
      \ifnum \i = \nint
        \pgfmathsetmacro\last{int(\outidx + 1)}
        \draw[->] (e-\outidx) -- (e-\last);
        \draw[->] (q-\outidx) -- (q-\last);
      \fi
      \pgfmathsetmacro\outidx{int(\nint + 1)}
      \pgfmathsetmacro\last{int(\nint + 2)}
      \node[neuron=NNYellow, right=10mm of q-\last] (mu) {\color{black}$\tilde{\boldsymbol{\mu}}_i$};
      \node[neuron=NNYellow, left=10mm of q-\last] (nacr) {\color{black}$\tilde{\boldsymbol{d}}_{ij}$};
      \draw[->] (q-\last) -- (mu);
      \draw[->] (q-\last) -- (nacr);
    }
  \end{tikzpicture}}
  \caption{
  Schematic illustration of the hierarchically interacting particle neural network (HIP-NN) structure. The blue squares are the input tensors of the network, consisting of the species ($Z$) and coordinate ($R$) of each atom $A_k$. The green rectangles represent the interaction layers which transmit information between neighboring atoms within a cutoff distance $R_{cut}$. The red squares represent the on-site (atom) layers, where the features of a specific atom are processed. 
  Black squares represent the intermediate outputs of the network. The final outputs of the network, shown in yellow squares, are the energies ($\tilde{E}_i$), transition dipoles ($\tilde{\boldsymbol{\mu}}_i$), and the non-adiabatic coupling vectors ($\tilde{\boldsymbol{d}}_{ij}$), where $i$ and $j$ are the indices of the states and the ``tilde'' denotes predicted values. For example, the total molecular energy $\tilde{E}_i$ , includes contributions $\tilde{E}_{k,i}^n$ at all atoms $A_k$ and hierarchical levels $n$. Note that the effective interaction length-scale between atoms grows linearly with respect to $n$. }
  \label{fig:hipnn}
\end{figure}